\documentclass[twocolumn,showpacs,pra,aps]{revtex4}
\usepackage{array}
\usepackage{amsmath}
\usepackage{graphicx}
\usepackage{amstext}
\usepackage{psfig}
\usepackage{amsfonts}
\begin{document}

\title{Quantum computation using weak nonlinearities:
robustness against decoherence}
  
\author{Hyunseok Jeong}

\email{jeong@physics.uq.edu.au}

\affiliation{Centre for Quantum Computer Technology, 
Department of Physics, University of Queensland, St
Lucia, Qld 4072, Australia}

\date{\today}

\begin{abstract}
We investigate decoherence effects in 
the recently suggested
quantum computation scheme using weak nonlinearities,
strong probe coherent fields,
detection and feedforward methods.
It is shown that in the weak-nonlinearity-based quantum gates,
decoherence in nonlinear media
{\it can be made arbitrarily small} simply by using
arbitrarily strong probe fields,
if photon number resolving detection is used.
On the contrary, we find that homodyne detection with feedforward
is not appropriate for this scheme
because in this case decoherence rapidly increases 
as the probe field gets larger. 
\end{abstract}


\maketitle

\section{Introduction}
Decoherence \cite{Zurek91} is
one of the main obstacles to 
the observation of
 quantum phenomena and
the realization of quantum information processing (QIP).
Since it is impossible to perfectly isolate a quantum system from its
environment, decoherence effects are more or less unavoidable.
Long-term existence of 
a macroscopic quantum superposition
\cite{Schr}
is hindered by the decoherence effects \cite{Zurek91,Kim92}.
Overcoming the destructive effects of decoherence is the central issue
for the realization of a large scale quantum computation (QC).
Quantum error correcting codes 
and entanglement purification protocols
have been developed to overcome the destructive effects
of decoherence \cite{Bennett96}.

Strong nonlinear effects
in optical systems, on the other hand, could be very useful for 
the observation of 
quantum phenomena \cite{Yurke,Tu,MB} and
the implementation of optical
QIP \cite{NC}.
Since currently available nonlinearities are extremely weak,
optical fields need to 
pass through long nonlinear media
for observable realizations of quantum effects.
This would cause the decoherence effects overwhelming so that
no quantum effects can actually manifest.

Recently, the idea of using weak cross-Kerr nonlinearities combined
with strong coherent fields has been developed 
by several different authors
and applied to various applications \cite{Frs,MunroQND,JeongPhD,Jeong05,
Barrett05,NM04,update1,update2,NM05pla,Jeong04q,Kim05}. 
The general idea of the weak-nonlinearity-based approach
is that the weak strength of a nonlinearity
can be compensated by using a strong probe coherent field,
$|\alpha\rangle$, with a very
large amplitude $\alpha$.
In particular,
Nemoto and Munro suggested a QC scheme
using weak nonlinearities and linear optics \cite{NM04}, which
has been further developed by Munro {\it et al.} \cite{update1,update2}.
They also pointed out \cite{NM05pla} that 
the weak-nonlinearity-based QC \cite{NM04,update1,update2} 
has merit over the linear optics QC
based on Knill {\it et al.}'s proposal \cite{KLM}
for a large scale quantum computation.
However, a rigorous investigation of
decoherence effects is essential to
verify the validity of the weak-nonlinearity-based
QC in a real experiment. 

Very recently, 
it was shown  \cite{Jeong05}
that a generation scheme for macroscopic superposition states \cite{Gerry} 
combined with the weak-nonlinearity-based approach 
\cite{Frs,MunroQND,Jeong05,JeongPhD,Barrett05,NM04}
 can {\it per se} overcome decoherence in the nonlinear medium,
 i.e., as the amplitude $\alpha$ becomes large, decoherence
during the nonlinear interaction {\it decreases}.  
In the concluding remarks of Ref.~\cite{Jeong05},
it was naively conjectured that 
the QC scheme
\cite{NM04,update1,update2,NM05pla} could also overcome the decoherence effects
in the same way.
However, it is unclear whether the weak-nonlinearity-based QC
can truly overcome decoherence during the nonlinear interactions
in this way with different detection-feedforward strategies 
\cite{NM04,update1,update2,NM05pla}.

In this paper, we
investigate decoherence effects in the weak-nonlinearity-based QC
with homodyne detection \cite{NM04,update1,NM05pla} and photon
number resolving detection \cite{update2}. 
We show that
as the initial amplitude of the probe coherent state gets larger,
decoherence rapidly {\it increases} in 
the two-qubit parity gate
with homodyne detection 
\cite{Barrett05,NM04,update1,NM05pla}.
On the contrary, we find that 
as the initial amplitude of the probe coherent state gets larger,
decoherence {\it diminishes} in
the two-qubit parity gate with photon number resolving 
detection \cite{update2}.
In other words, decoherence
can be made {\it arbitrarily small} in this type of gates simply by
increasing the probe field amplitude.
We explain that this is due to the difference of the geometric requirements
in the phase space.
Since the two-qubit parity gate is the key element in
the weak-nonlinearity-based QC \cite{NM04,update1,update2,NM05pla},
our result shows that the
weak-nonlinearity-based QC can naturally overcome decoherence effects
but photon number resolving 
detection is needed for such robustness to decoherence.

\begin{figure}
\centerline{\scalebox{0.55}{\includegraphics{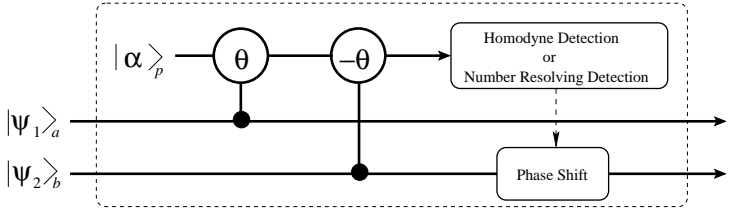}}}
 \caption{A schematic of the parity gate using weak nonlinearities
 that entangles two qubits.
The two-mode nonlinear interactions $\theta$ and $-\theta$ occur
only for the horizontally polarized qubit state $|H\rangle$. 
 } \label{fig:gate}
\end{figure}
 
\section{
The two-qubit parity gate using weak nonlinearities}
It was shown that in conjunction with a strong coherent field,
two weak nonlinearities are sufficient to implement a 
parity gate that entangles two qubits
as illustrated in Fig.~\ref{fig:gate} \cite{NM04,Barrett05}.
The interaction Hamiltonian of the cross-Kerr nonlinearity 
between modes $a$ and $p$ is
$H_{K}=\hbar \chi \hat a^\dagger_a \hat a_a \hat a_p^\dagger \hat a_p$,
where $\hat a$ ($\hat a^\dagger$) represents the annihilation (creation) operator
and $\chi$ is the nonlinear coupling constant.
The interaction between a Fock state, $|n\rangle_a$,
and a probe coherent state, $|\alpha\rangle_p$,
is described as
$U_K(t)|n\rangle_a|\alpha\rangle_p
=|n\rangle_a|\alpha e^{in\theta}\rangle_p$,
where $\theta=\chi t$ with the interaction time $t$,
and $U_K(t)=e^{i H_K t/\hbar}$.
Using polarization beam splitters, 
it is possible to use 
the horizontally and  vertically polarized single-photon states,
$|H\rangle$ and  $|V\rangle$, 
to work as \cite{NM04}
$U_K(t)|H\rangle|\alpha\rangle=|H\rangle|\alpha e^{i\theta}\rangle$ and
$U_K(t)|V\rangle|\alpha\rangle=|V\rangle|\alpha\rangle$.
For simplicity, we assume two identical initial qubits,
$|\Psi\rangle_a=(|H\rangle_a+|V\rangle_a)/\sqrt{2}$ and
$|\Psi\rangle_b=(|H\rangle_b+|V\rangle_b)/\sqrt{2}$.
The total initial state is
\begin{equation}
|\psi_i\rangle=\frac{1}{2}
\big(|H\rangle+|V\rangle\big)_a
\big(|H\rangle+|V\rangle\big)_b|\alpha\rangle_p
\label{initial}
\end{equation}
where $\alpha$ is assumed to be real without losing generality.
After the first nonlinear interaction 
between modes $a$ and $p$
with angle $\theta$, the initial state evolves to
$|\psi_1\rangle=
\{\big(|HH\rangle
+|HV\rangle\big)_{ab}|\alpha e^{i\theta}\rangle_p
+\big(|VH\rangle
+|VV\rangle\big)_{ab}|\alpha\rangle_p\}/2$.
After the second nonlinear interaction
between modes $b$ and $p$
with angle $-\theta$, it becomes
\begin{equation}
\begin{aligned}
|\psi_2\rangle=&
\frac{1}{2}\Big\{\big(|HH\rangle
+|VV\rangle\big)_{ab}|\alpha\rangle_p\\
&+|HV\rangle_{ab}|\alpha e^{i\theta}\rangle_p
+|VH\rangle_{ab}|\alpha e^{-i\theta}\rangle_p\Big\}.
\end{aligned}
\end{equation}
A measurement is then performed 
to distinguish the probe beam $|\alpha\rangle_p$
from $|\alpha e^{i\theta}\rangle_p$ 
and $|\alpha e^{-i\theta}\rangle_p$, while it 
does not distinguish $|\alpha e^{i\theta}\rangle_p$ 
and $|\alpha e^{-i\theta}\rangle_p$.

\begin{figure}
\centerline{(a)}
\centerline{\scalebox{0.4}{\includegraphics{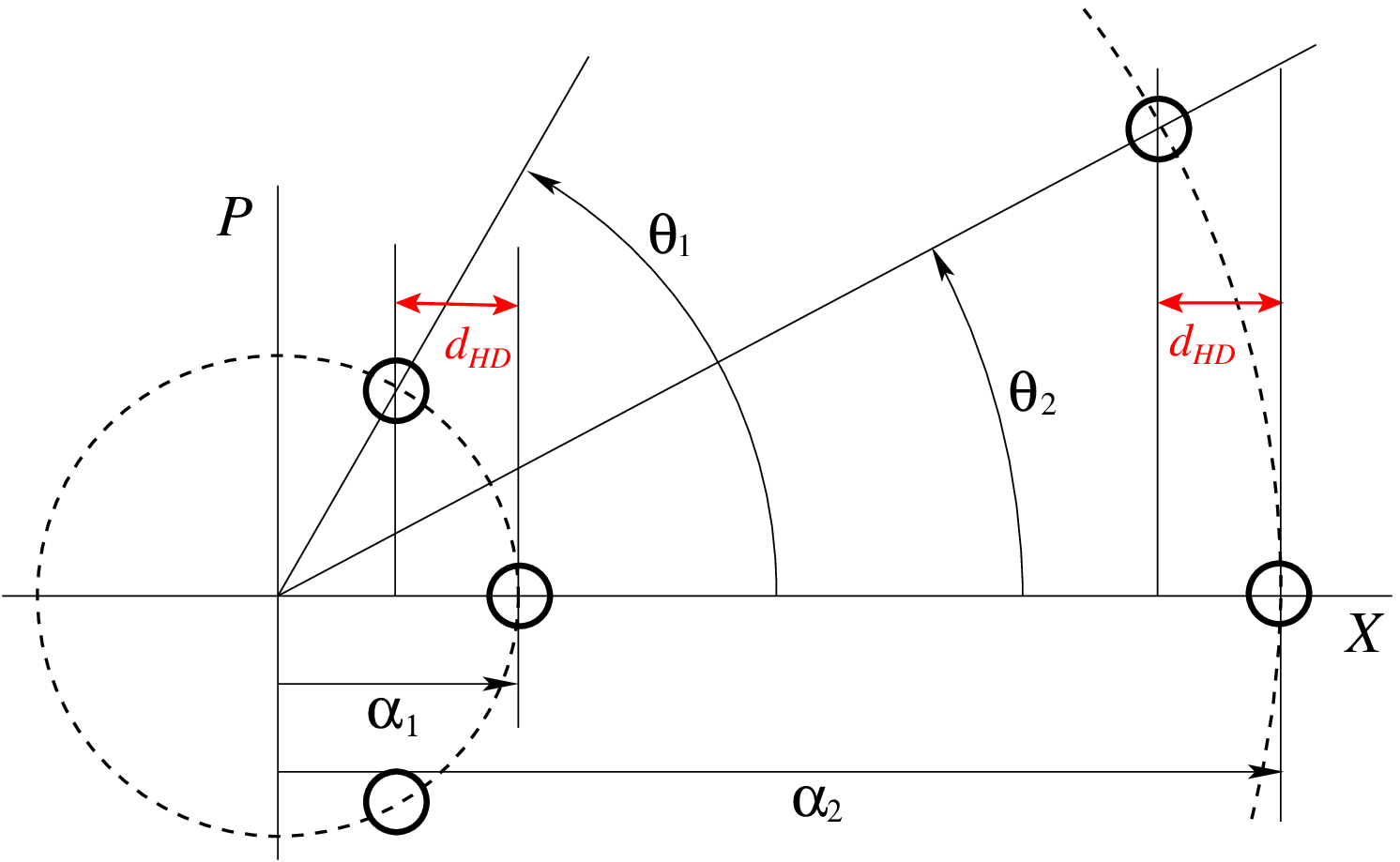}}}
\vspace{0.3cm}
\centerline{(b)}
\centerline{\scalebox{0.4}{\includegraphics{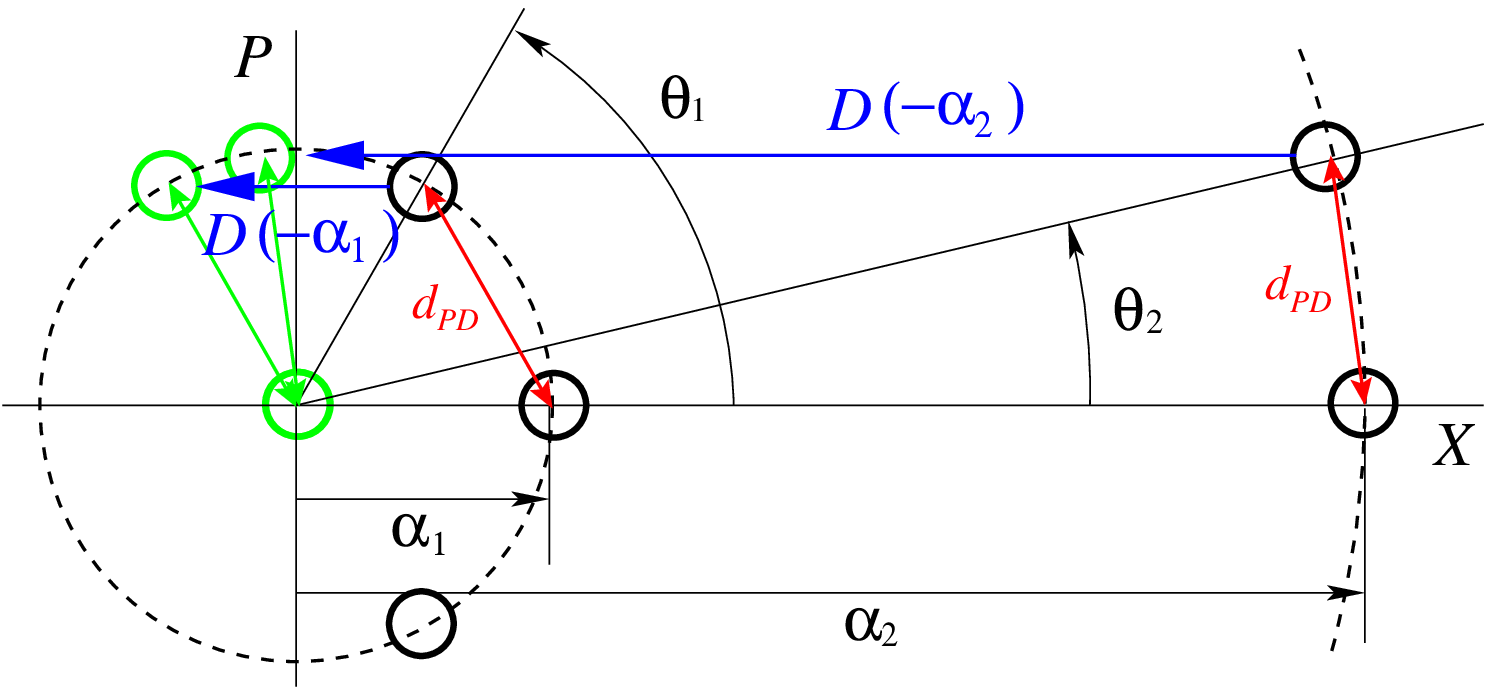}}}
 \caption{Geometric diagram for the weak-nonlinearity-based 
two-qubit parity gate using (a) homodyne detection
and (b) photon number resolving detection.
(a) As the initial amplitude becomes large, the ``travel path'',
$\alpha\theta$,
of the coherent state in the phase space
should increase
to maintain the distance $d_{HD}$, i.e.,
 $\alpha_2\theta_2>\alpha_1\theta_1$.
This causes the increase of decoherence effects for large amplitudes.
(b) Regardless of the initial amplitude, the travel path,
$\alpha\theta$, of the coherent state of the same order
can maintain the distance $d_{PD}$, i.e.,
$\alpha_2\theta_2\approx\alpha_1\theta_1$. 
This reduces the decoherence effects because the interaction time
$t$ in a nonlinear medium becomes shorter
as the initial amplitude increases.} \label{fig:compare}
\end{figure}

Suppose that homodyne detection
for quadrature $\hat{X}=(a+a^\dagger)/2$ 
is performed with the measurement result $X$.
As can be seen in Fig.~\ref{fig:compare}(a),
the distinguishability of the measurement is determined by
the distance
\begin{equation}
d_{HD}=\alpha(1-\cos\theta)\approx\frac{\alpha\theta^2}{2} 
\label{eq:t1}
\end{equation}
where the approximation has been made under the assumptions of
$\theta\ll1$. Munro {\it et al.} pointed out that the error probability is
$P_{err}\approx 10^{-4}$ for $d_{HD}=4$ \cite{update1}.
If $d_{HD}$ is large enough and
the measurement outcome is $X>X_{mid}$,
where $X_{mid}=\alpha(1+\cos\theta)/2$, the output state is
\begin{equation}
|\psi_f\rangle=\frac{1}{\sqrt{2}}\big(|HH\rangle+|VV\rangle\big)_{ab}.
\label{hhvv}
\end{equation}
On the other hand, if $X<X_{mid}$, the output state is
$|\psi_f^\prime\rangle=
\big(e^{i\phi(X)}|HV\rangle+e^{-i\phi(X)}|VH\rangle\big)_{ab}/\sqrt{2}$,
where $\phi(X)=2\alpha\sin\theta(X-\alpha \cos \theta)$.
One can transform the state $|\psi_f^\prime\rangle$
to the state $|\psi_f\rangle$ by
a simple phase shift for mode $b$ \cite{NM04,update1}
based upon the measurement result.

Instead of homodyne detection, the photon number resolving
detection can be used 
to distinguish the coherent state elements for mode $p$
as shown in Fig.~\ref{fig:compare}(b). 
\cite{update2}.
The displacement operation $D(-\alpha)$
is then applied 
before photon number detection is performed.
After the displacement operation, the state $|0\rangle$ can be distinguished from
$|\alpha(e^{\pm\theta}-1)\rangle$ by photon number resolving
detection with measurement $n_p$.
The output state (\ref{hhvv}) is obtained for $n_p=0$.
The resulting state for $n_p\neq 0$,
$\big(e^{i\phi(n_p)}|HV\rangle+e^{-i\phi(n_p)}|VH\rangle\big)_{ab}/\sqrt{2}$,
can be transformed to the state (\ref{hhvv})
by a phase shift with the phase factor
$\phi(n_p)=n_p\tan^{-1}[\cot(\theta/2)]$.
The distance $d_{PD}$ which determines the distinguishability of
the measurement is
\begin{equation}
d_{PD}=2\alpha\sin\frac{\theta}{2}\approx\alpha\theta
\label{eq:t2}
\end{equation}
and the error probability
is $P_{err}\approx 10^{-4}$ for $d_{PD}=\pi$ \cite{update2}.

\section{Decoherence in the weak-nonlinearity-based 
parity gate}
The ideal output state of the two-qubit parity gate
should be the pure entangled state (\ref{hhvv}).
However, the actual outcome state will be in
a mixed state due to the decoherence effects {\it in
the nonlinear media}.
Photon losses may occur both in the probe field mode ($p$)
and in the qubit modes ($a$ and $b$).
However, the possibility 
of losing photons in the qubit modes becomes
lower as the initial amplitude gets larger, because the interaction times
($t=\theta/\chi$)
in the nonlinear media
become shorter as can be shown from Eqs.~(\ref{eq:t1}) and (\ref{eq:t2}).
The important factor of decoherence
in the two-qubit output state
(\ref{hhvv})
 is photon losses in the probe field mode.
Since the coherent field contains a large number of photons, 
it is easy to lose photons even in a very short time. Such photon losses
in the probe coherent field cause the loss of phase information in the 
two-qubit output state
(\ref{hhvv}).
In particular, 
it is known that a superposition of two distant coherent states
rapidly loses its coherence even when it loses a small number of photons
\cite{Kim92}.
Therefore, photon losses of the probe coherent field 
in the nonlinear media should be considered 
the main source of decoherence in the two-qubit output state. In what follows,
we shall consider photon losses in the probe field and decoherence effects
in the two-qubit output state caused by such photon losses.

Suppose that the probe coherent field loses photons
in the first nonlinear medium 
as $|\alpha\rangle\rightarrow|{\cal A}\alpha\rangle$,
where we define the amplitude parameter ${\cal A}~(\leq 1)$.
After the first nonlinear interaction, the total initial
state becomes a mixed state as
\begin{eqnarray}
&&
\frac{1}{4}\Big\{
\big(|HH\rangle+|HV\rangle\big)
\big(\langle HH|+\langle HV|\big)\otimes
|{\cal A}\alpha e^{i\theta}\rangle\langle{\cal A}\alpha e^{i\theta}|
\nonumber\\
&&+{\cal C}\big(|HH\rangle+|HV\rangle \big)\big(\langle VH|+\langle VV|\big)
\otimes|{\cal A}\alpha e^{i\theta}\rangle\langle{\cal A}\alpha |\nonumber\\
&&+{\cal C}^*\big(|VH\rangle+|VV\rangle\big)\big(\langle HH|+\langle HV|\big)
\otimes|{\cal A}\alpha \rangle\langle{\cal A} \alpha e^{i\theta}|\nonumber\\
&&+\big(|VH\rangle+|VV\rangle\big)\big(\langle VH|+\langle VV|\big)
\otimes|{\cal A}\alpha \rangle\langle{\cal A}\alpha |\Big\}_{abp}
\end{eqnarray}
where the coherent parameter, $\cal C$, is introduced 
to quantify the degree of dephasing.
It is easy to recognize that {\it both} $\cal A$ and $\cal C$
should be reasonably large for the two-qubit parity gate to work properly
at the end.
If $\cal A$ is large but $\cal C$ is negligible, the final output state
of the parity gate will be
$\rho^{(m)}_f=
(|HH\rangle\langle HH|+|VV\rangle\langle VV|)_{ab}/2$, which was
also pointed out in Ref.~\cite{update2}.
This completely dephased state, $\rho^{(m)}_f$, does not contain entanglement,
i.e.,
the two-qubit parity gate completely fails. 
On the other hand, 
if $\cal C$ is close to 1 but $\cal A$ is negligible, the final result will be
$|\phi_f\rangle=
\big(|H\rangle+|V\rangle\big)_a
\big(|H\rangle+|V\rangle\big)_b/2$,
which is simply identical to the unentangled initial qubits so that the 
gate also fails.

The decoherence effects for a state described by
the density operator $\rho$
can be induced by solving the master equation \cite{Phoenix}
\begin{equation}
\label{master-eq}
  {\partial \rho \over \partial t}=\hat{J}\rho +\hat{L}\rho~;~\hat{J}\rho=\gamma a\rho a^\dag,~~
  \hat{L}\rho=-{\gamma \over 2}(a^\dag a\rho +\rho a^\dag a)
\end{equation}
where $\gamma$ is the energy decay rate.
The formal solution of the
master equation (\ref{master-eq}) can be written as
$\rho(t)=\exp[(\hat{J}+\hat{L})t]\rho(0)$
where $t$ is the interaction time.
The evolution of the initial density element
$|\alpha\rangle\langle\beta|$
by the decoherence process 
$\tilde{\cal D}$ 
can be described as \cite{Phoenix}
\begin{equation}
{\tilde{\cal D}}(|\alpha\rangle\langle\beta|)=
e^{-(1-e^{-\gamma t})\{\frac{1}{2}(|\alpha|^2+|\beta|^2)-\alpha\beta^*\}} 
  |{\cal A} \alpha \rangle\langle {\cal A}\beta |,
\label{solution-master}
\end{equation}
where $|\beta\rangle$ ($|\alpha\rangle$) is a coherent state with
amplitude $\beta$ ($\alpha$) and ${\cal A}=e^{-\gamma t/2}$.  
However, it should be noted that the decoherence process 
($\cal\tilde D$) occurs {\it simultaneously} 
with the unitary evolution ($\cal\tilde U$) by the cross-Kerr interaction
Hamiltonian $H_K$ in a nonlinear medium. 
This combined  process 
can be modeled as follows \cite{Jeong05}.
One may assume that $\cal\tilde U$ occurs for a short time
$\Delta t$, and then $\cal\tilde D$ occurs for another $\Delta t$.
In other words, $\cal\tilde U$ and $\cal\tilde D$
continuously take turn for such short intervals in the nonlinear medium.
By taking $\Delta t$ arbitrarily 
small, one can obtain an extremely
good approximation of this process
 for a given time $t$ ($=N\Delta t$) with large integer number $N$.
Let us set 
$\Delta\theta=\chi\Delta t=\pi/N$.
In our calculation, we have chosen $N=10^6$, i.e.,
$\Delta \theta=\pi/10^6$.
This value gives a very good approximation
for the whole range of $\alpha$ in our study
\cite{Jeong05}.
Using this model, let us fist consider the evolution of
one cross term, 
$(|HH\rangle\langle VH|)_{ab}\otimes(|\alpha\rangle\langle\alpha|)_p$, in
the initial state (\ref{initial}).
After time $t$ $(=N\Delta t)$ in the nonlinear medium, it evolves to
\begin{equation}
\begin{aligned}
&\Big\{{\cal\tilde D}_p(\Delta t){\cal\tilde U}_{ap}(\Delta t)\Big\}^N
\big(|HH\rangle\langle VH|\big)_{ab}\otimes(|\alpha\rangle\langle\alpha|)_p\\
&~~~~~={\cal C}\big(|HH\rangle\langle VH|\big)_{ab}\otimes
(|{\cal A}\alpha e^{i\theta}\rangle\langle {\cal A}\alpha|)_p
\end{aligned}
\end{equation}
where ${\cal\tilde U}(\Delta t)\rho\equiv U_K(\Delta t)\rho U_K^\dagger(\Delta t)$
and
\begin{equation}
\begin{aligned}
&{\cal C}=\exp\Big[-\alpha^2(1-e^{-\gamma(t/N)})\\
&~~~~~\sum_{n=1}^N\exp[-\gamma(t/N)]^{(n-1)}(1-\exp[-i\chi n(t/N)])\Big].
\end{aligned}
\end{equation}
The amplitude parameter $\cal A$ and the
coherence parameter $\cal C$
can then be obtained for an initial amplitude $\alpha$.
We shall use the absolute value of the coherence parameter
$|{\cal C}|$ to assess the degree of dephasing.

\begin{figure}
\centerline{(a)}
\centerline{\scalebox{0.65}{\includegraphics{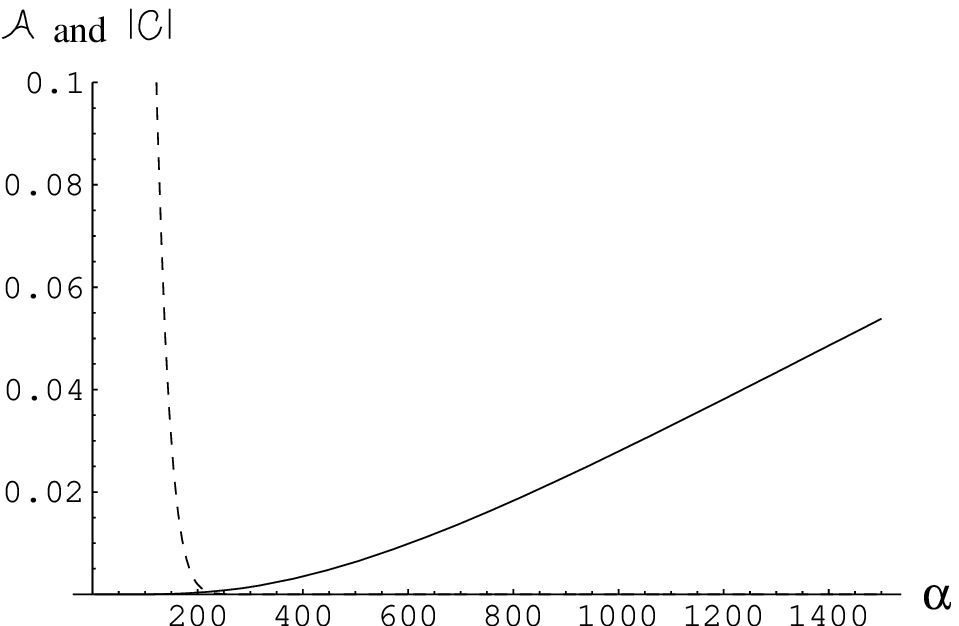}}}
\vspace{0.2cm}
\centerline{(b)}
\centerline{\scalebox{0.65}{\includegraphics{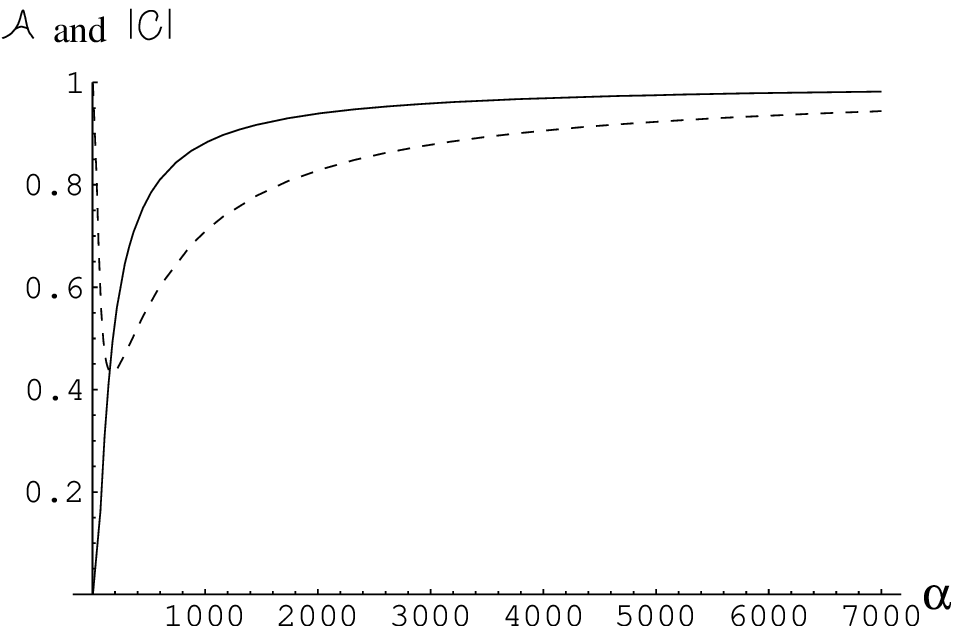}}}
 \caption{The amplitude parameter $\cal A$ (solid line)
and the absolute coherence parameter $|{\cal C}|$ (dashed line)
against the initial amplitude $\alpha$
for homodyne detection and photon number detection.
The two-qubit parity gate works when both $\cal A$ and $|\cal C|$
are large. 
(a) When homodyne detection is used:
it is obvious that the condition
of both $\cal A$ and $|\cal C|$ being large cannot be met.
(b) When photon number resolving detection is used:
the condition is satisfied for a large $\alpha$. 
 } \label{fig:AC}
\end{figure}

We are interested in $\cal A$ and $|{\cal C}|$ 
under experimentally realistic assumptions.
It has been known that an optical fiber of about $3,000km$
may be required for a nonlinear interaction of $\theta=\pi$
using a currently available cross Kerr nonlinearity
\cite{SandersMilburn}.
We first choose $\chi/\gamma=0.0125$ 
that the amplitude will reduce as ${\cal A}\approx0.533$ for $15km$ while 
$\theta=\pi$ is obtained for $3,000km$.
This corresponds 0.364dB/km of signal loss, 
which is a typical value for commercial fibers used
for telecommunication and easily achieved using current technology
\cite{fiberloss1,fiberloss2}.
Note that signal losses in some pure silica core fibers are even less than 0.15dB/km 
\cite{fiberloss2}.
The Fig.~\ref{fig:AC}(a) shows 
that as the initial amplitude $\alpha$ increases for a fixed $d_{HD}(=4)$,
the absolute coherence parameter $|{\cal C}|$ (dashed line) rapidly decreases
for the homodyne detection scheme.
The absolute coherence parameter $|{\cal C}|$ is not negligible only when $\alpha$ is small.
However, the two-qubit parity gate does not work in this regime because
$\cal A$ (solid line) becomes extremely small.
This means the probe coherent state becomes the pure vacuum so that
a large $|\cal C|$ is meaningless.
The Fig.~\ref{fig:AC}(b) shows that this
scheme with photon number detection 
does not suffer such problems:
as the initial amplitude $\alpha$ increases for a fixed $d_{PD}(=\pi)$,
both $\cal A$ and $|{\cal C}|$ increases for large $\alpha$.
Some detailed values 
for $\chi/\gamma=0.0125$ (0.364dB/km) and $\chi/\gamma=0.0303$ (0.15dB/km)
including the required length of optical fibers
have been presented in Table~\ref{table1}.

\begin{table}
\caption{\label{table1}
The amplitude parameter $\cal A$ and coherence parameter $|\cal C|$
under various conditions. $\alpha$ is the initial amplitude of the probe
coherent state and `Length' is the required length of the optical fiber.
(a) Cases for homodyne detection with $d_{HD}(\approx\alpha\theta^2/2)=4$.
Comparing $\cal A$ and $|{\cal C}|$, it is obvious that this detection strategy
cannot be used for the weak-nonlinearity-based two-qubit parity gate.
(b) Cases for photon number resolving detection
with $d_{PD}(\approx\alpha\theta)=\pi$. Both $\cal A$ and $|{\cal C}|$
approach 1 simultaneously when $\alpha$ becomes large.}
\begin{ruledtabular}
\centerline{(a) Homodyne detection}
\begin{tabular}{ccccccc}
 $\chi/\gamma$  & $\theta(=\chi t)$
   &  $\alpha$
     & Length (km)  & $\cal A$    & $|{\cal C}|$   \\
\hline
 & 0.284    &  100   &  271 &   $10^{-5}$    &      0.210 \\
$0.0125$   & 0.163    &  300   & 130 &   $0.0014$    &     $ \sim0$
  \\
   & 0.052    &  3000 &     50 &  $0.127$   &     $ \sim0$   \\
\hline
  & 0.284    &  100    &   271 &  0.009    &      $10^{-4}$ \\
 $0.0303$   & 0.163    &  300   &    130 & $0.067$ & $ \sim0$ \\
  & 0.052    &  3000    &  50 &  0.427   &     $ \sim0$   
\end{tabular}
\end{ruledtabular}
\vspace{0.1cm}
\begin{ruledtabular}
\centerline{(b) Photon number resolving detection}
\begin{tabular}{ccccccc}
 $\chi/\gamma$  & $\theta(=\chi t)$
   &  $\alpha$
     & Length (km)  & $\cal A$    & $|{\cal C}|$   \\
	 \hline
 & 0.0105 & 300 &   10 & 0.658 & 0.474 \\
 0.0125 & $1.05\times10^{-3}$  & 3000 &    1 & 0.959  & 0.878 \\
 & $1.05\times10^{-4}$ & $3\times 10^{4}$  &   0.1 & 0.996 & 0.985 \\
\hline
    & 0.0105 & 300 &   10    &  0.841   &    0.644   \\
 0.0303 & $1.05\times10^{-3}$  & 3000 &  1  & 0.983 & 0.946 \\
   & $1.05\times10^{-4}$ & $3\times 10^{4}$  &  0.1  &  0.998   &  $>0.99$ 
\end{tabular}
\end{ruledtabular}
\end{table}

One can understand the difference between 
Fig.~\ref{fig:AC}(a) and Fig.~\ref{fig:AC}(b)
 by a simple geometric analysis in Fig.~\ref{fig:compare}.
For the case of homodyne detection,
as the initial amplitude $\alpha$ gets larger, the ``travel path'' of
the coherent state in the phase space, $\alpha\theta$,
increases.
 (Even though  $\theta$ decreases, the increase of $\alpha$ makes
$\alpha\theta$ larger for a fixed $d_{HD}(\approx\alpha\theta^2/2)$
as shown in Fig.~\ref{fig:compare}(a).)
 Therefore, the initial coherent state should travel
longer in the phase space.
This makes decoherence actually increase as $\alpha$ gets larger.
In this case, the 
principle of increasing $\alpha$ to compensate small $\theta$
will not work efficiently.
However, the mechanism is totally different
when photon number resolving detection is used.
As the initial amplitude $\alpha$ gets larger, the travel path $\alpha\theta$
does not increase for a fixed $d_{PD}(\approx\alpha\theta)$ 
but remains approximately the same 
(see Fig.~\ref{fig:compare}(b)).
Therefore, the coherent state travels the same distance regardless of
$\alpha$, while the interaction time $t(=\theta/\chi)$ depending on $\theta$
keep decreasing as $\alpha$ increases.
Such decrease of the interaction time $t$
for the same distance causes the decrease of decoherence.

\begin{figure}
\centerline{\scalebox{0.65}{\includegraphics{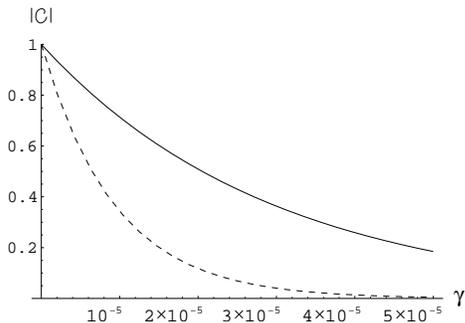}}}
 \caption{
The absolute coherence parameter $|{\cal C}|$
against the energy decay rate $\gamma$.
The nonlinear strength is assumed to be $\chi=0.01$.
Note that the amplitude parameter $\cal A$ will be
always close to 1 in this regime since $\gamma t$
is very small. 
The solid line corresponds to $\alpha=10^3$
so that $\theta_{HD}=\chi t\approx0.13$.
The dashed line 
corresponds to $\alpha=10^4$
and $\theta_{HD}=\chi t\approx0.04$.
The coherence parameter $|{\cal C}|$ rapidly decreases
for a small increase of $\gamma$.
 } \label{fig:rev}
\end{figure}

So far, we have considered optical fibers,
in which the energy decay rate $\gamma$ is typically
much larger than the nonlinear strength $\chi$
as shown in Table I.
Munro {\it et al.} 
discussed quantum non-demolition (QND) measurements
using giant cross-Kerr nonlinearities available in
electromagnetically induced transparency (EIT)
\cite{MunroQND}.
This technique can be used to avoid the problems of highly absorptive media.
Using this technique,
the weak-nonlinearity based QC can be 
performed in the regime of 
$\gamma t \approx \chi t(=\theta)\ll1$.
This change will certainly improve $\cal A$ and $|{\cal C}|$
for a given value of $\theta$ (or a given value of $\alpha$)
in the case of photon number resolving detection.
For example, if $\chi=0.01$ and $\gamma=0.01$,
${\cal A}\approx0.999$ and  $|{\cal C}|\approx0.98$ are obtained
for $\theta\approx0.0105$
(i.e. $\alpha=300$).
However, the realization of the weak-nonlinearity based QC
using homodyne detection
will still be extremely hard.
We set $\chi=0.01$ and consider the following two examples
in Fig.~4.
In the first example, $\alpha=10^3$ where the required angle is
$\theta\approx0.13$, and in the second,
$\alpha=10^4$ and $\theta\approx0.04$.
Fig.~4 clearly shows that
the energy decay rate $\gamma$ should be extremely small 
in the case of homodyne detection, which is
not realistic using current technology.

\section{Discussions}
In the weak-nonlinearity-based QC scheme,
a strong probe coherent field
with a large amplitude is necessarily required.
However, as the amplitude of the coherent field
gets larger, decoherence during a nonlinear interaction
rapidly {\it increases} 
when homodyne detection is used.
On the contrary,
decoherence {\it diminishes}  under the same condition
when the photon number resolving
measurement is used.
This shows that
the weak-nonlinearity-based QC
can naturally overcome decoherence during the nonlinear interactions
simply by using strong probe fields, when photon number resolving detection
is used.

Since $d_{PD}=\pi$ in Eq.~(\ref{eq:t2})
is required for a small error probability,
the photodetector for the two-qubit gate 
should be able to discriminate 
about 10 ($\approx d_{PD}^2$) photons.
Such detection ability is extremely demanding
using current technology.
It may be crucial to first
develop the photon number resolving QND technique
using a weak nonlinearity,
a strong coherent field and homodyne detection in Ref.~\cite{MunroQND},
which was employed for the two qubit gate in Ref.~\cite{update2}.
Here, we point out that the QND technique
in Ref.~\cite{MunroQND} does {\it not} suffer the increase
of decoherence
for large probe field amplitudes 
in the nonlinear medium 
because distinguishability of this QND 
scheme \cite{MunroQND} depends on
$\approx\alpha\theta$,
not on $\approx\alpha\theta^2$.

Using photon number resolving detection 
in the weak-nonlinearity-based two-qubit parity gate
also requires a highly precise displacement operation, $D(-\alpha)$,
with a very large $\alpha$.
The displacement operation can be performed using
a strong coherent field
and a beam splitter with high transmittivity. 
It would be experimentally challenging since the average
photon number of the probe coherent field 
should be $|{\cal A}^2\alpha|^2\gg 10^6$ to obtain good coherence
as can be seen in Table I(b).

The two-qubit parity gate which we have considered
in this paper
is based on two weak nonlinearities, a probe coherent field,
a probe beam measurement 
and classical feedforward 
as shown in Fig.~1
\cite{NM04,update1,update2,NM05pla}. 
Here, we note the recently suggested
weak-nonlinearity-based
controlled-phase gate by Spiller {\it et al.} \cite{Spiller},
where the probe beam measurement is not necessary at the cost of
using additional nonlinearities and displacement operations.
The requirement
of successful achievement of this gate is $\alpha\theta\sim1$
\cite{Spiller,private}, which satisfies
the condition for the robustness against decoherence analyzed
in this paper.
Therefore, such an approach without the probe beam measurement
may be considered as an alternative of the weak-nonlinearity-based
parity gate in Fig.~1 using photon number resolving detection.

We have explained the reason for the decrease of decoherence
when photon number resolving detection is used. Decoherence depends on
the interaction time $t$ and the distance $d$ in Fig.~2.
In other words, decoherence
increases when either $t$ or $d$ increases.
As the initial amplitude $\alpha$ of the probe beam increases,
the interaction time $t$ is reduced while
$d$ remains the same order when 
photon number resolving detection is used, which causes the decrease of
decoherence. 
This is {\it not} the case for 
homodyne detection for which $d$ should increase as
$\alpha$ increases, which results in the increase of decoherence.

We finally point out that there are certain types errors
which cannot be made small
in the way discussed in this paper.
For example, in a real experiment, self-phase modulation (SPM)
will also occur during the cross-Kerr interactions in optical fibers
\cite{agra01}.
It may hinder measuring phase shift purely induced by the cross-Kerr effects
in the weak-nonlinearity based QC. 
Note that the principle of the weak-nonlinearity based approach is to
use a large-amplitude probe beam to compensate the weak strength of
the nonlinearity. This mechanism is applied to
both the cross-Kerr effects and the SPM effects \cite{Jeong04q}. 
Therefore, the {\it ratio} between the cross-Kerr effect
and the SPM effect during the interaction between
the probe beam and the single-photon qubit in the nonlinear medium
will {\it not} be affected by the detection strategy
or the initial amplitude.
Therefore, this kind of errors 
should be separately  dealt with (for example, see Ref.~\cite{Li}).

This work was supported by the Australian
Research Council. The author would like to  
thank S.D. Barrett, T.C. Ralph, A.M. Branczyk and W.J. Munro 
for valuable comments and stimulating discussions.

\end{document}